\documentclass{article}
\usepackage{graphicx} 
\usepackage{graphicx,amsmath,amssymb,tikz,amsthm} 
\usepackage{arydshln}
\usepackage{mathtools}
\usepackage{bm}
\usepackage{cite}
\usepackage[linktocpage=true,
  colorlinks=true, 
  pdfborder={0 0 0},
  linkcolor=blue,
  citecolor=red,
  filecolor=yellow,
  urlcolor=blue,
  bookmarks,
  pdfauthor={},
]{hyperref}
\usetikzlibrary{quantikz2}
\usepackage{authblk}
\makeatletter
\renewcommand*\env@matrix[1][\arraystretch]{%
  \edef\arraystretch{#1}%
  \hskip -\arraycolsep
  \let\@ifnextchar\new@ifnextchar
  \array{*\c@MaxMatrixCols c}}
\makeatother

\newtheorem{theorem}{Theorem}
\newtheorem{problem}{Problem}
\newtheorem{proposition}{Proposition}
\newtheorem{corollary}{Corollary}

\newtheorem{remark}{Remark}

\title{An efficient quantum Hadamard product algorithm for functions}

\author[1,2]{Xinchi Huang\footnote{Email: kkou@quemix.com; huangxc@g.ecc.u-tokyo.ac.jp}}
\author[1,2]{Hirofumi Nishi}
\author[3]{Tomofumi Zushi}
\author[1,2,4]{Yu-ichiro Matsushita}
\affil[1]{Department of Physics, The University of Tokyo, Tokyo 113-0033, Japan}
\affil[2]{Quemix Inc., Taiyo Life Nihombashi Building, 2-11-2 Nihombashi Chuo-ku, Tokyo 103-0027, Japan}
\affil[3]{Sumitomo Rubber Industries, Ltd., 2-1-1 Tsutsui, Chuo, Kobe, Hyogo 651-0071, Japan}
\affil[4]{Quantum Materials and Applications Research Center, National Institutes for Quantum Science and Technology (QST), 2-12-1 Ookayama, Meguro-ku, Tokyo 152-8550, Japan}

\date{}

\begin{document}
\maketitle
\begin{abstract}
We propose an efficient quantum algorithm for preparing the Hadamard product state of two quantum states whose amplitudes are generated by functions on a uniform grid with grid number $N$. As the Hadamard product operation is non-unitary, the conventional approach generally suffer from a success probability that scales as $O(1/N)$, leading to an $O\left(\sqrt{N}\right)$ query complexity even with quantum amplitude amplification. 
Our method exploits the Fourier-space representation of the input functions, where the Hadamard product can be treated through a convolution structure and approximated using localized Fourier coefficients. The resulting quantum circuit has complexity governed by the Fourier regularity of the underlying functions rather than directly by the grid number. 
In particular, when either of the input functions has finitely many non-zero Fourier coefficients, the algorithm prepares the exact quantum Hadamard product state under $N$-independent query complexity. Moreover, we also propose a novel quantum circuit for the partial inner product as one of its applications. 
\end{abstract}

\section{Introduction and main results}
\label{sec1}

Quantum states whose amplitudes encode discretized functions naturally arise in quantum algorithms for numerical analysis, signal processing, and scientific computing. 
Under such settings, element-wise multiplication of amplitudes corresponds to the Hadamard product of underlying functions and is a fundamental and important nonlinear operation. 
However, implementing this operation on quantum devices is non-trivial because the Hadamard product is not unitary in general, and standard post-selection-based approaches (e.g., \cite{Holmes.2023,Ramezani.2023}) may have small success probability, which is inverse proportional to the grid number.

In this paper, we consider the quantum circuit for preparing the Hadamard product state for two quantum states, up to a given error bound, provided that their quantum state preparation oracles are known. 
If the input quantum states are given by the amplitude encoding of some underlying continuous functions, then we can employ the Fourier space representations to improve the success probability of post-selection. 

To state our main results, we begin with the mathematical settings and the description of our problem. Let $n\in \mathbb{N}$ and $N = 2^n$. Suppose two quantum states $\ket{\psi}$ and $\ket{\varphi}$ are given by
\begin{equation*}
\ket{\psi}_n = \sum_{j=0}^{N-1} \psi_j \ket{j}_n, \quad \ket{\varphi}_n = \sum_{j=0}^{N-1} \varphi_j \ket{j}_n.
\end{equation*}
In general, $\psi_j, \varphi_j$ can be complex-valued, while we mainly focus on the real-valued cases in Sects.~\ref{sec2}-\ref{sec3}. 
We define the quantum Hadamard product problem as follows:
\begin{problem}[QHP]
Given two oracles $U_\psi$ and $U_\varphi$ that prepares the quantum states $\ket{\psi}$ and $\ket{\varphi}$, respectively, as well as an error bound $\varepsilon>0$. 
Find a quantum circuit that prepares a quantum state $\ket{\phi_\mathrm{out}}$ which is $\varepsilon$-close to the following quantum state: 
\begin{align*}
\ket{\psi\varphi}_n = \frac{1}{\sqrt{\sum_{j=0}^{N-1}|\psi_j\varphi_j|^2}} \sum_{j=0}^{N-1} \psi_j\varphi_j \ket{j}_n,
\end{align*}
with at least a constant success probability (e.g., larger than $1/2$). Here, ``$\varepsilon$-close" means 
$$
\|\ket{\phi_{\mathrm{out}}}_n-\ket{\psi\varphi}_n\| \le \varepsilon,
$$
where $\|\cdot\|$ denote the $l^2$-norm of vectors. 
\end{problem}
The operation to solve the problem is not unitary, and hence, the success probability should be considered. Here, the word ``constant" means the probability is independent of $N$ and the quantum states $\ket{\psi}$, $\ket{\varphi}$. 
A conventional quantum circuit is known by using CNOT gates and measurements (i.e., post-selections). 
Then, we have the following proposition:
\begin{proposition}
\label{sec1:prop1}
There exists a quantum circuit to solve the QHP problem for arbitrary $\varepsilon\ge 0$ with $O\left(1/\sqrt{\sum_{j=0}^{N-1}|\psi_j\varphi_j|^2}\right)$ queries of the oracles $U_\psi$, $U_\varphi$ and $O\left(n^2/\sqrt{\sum_{j=0}^{N-1}|\psi_j\varphi_j|^2}\right)$ additional single-qubit or two-qubits gates.
\end{proposition}
The proposition can be proved directly by the conventional quantum circuit in Fig.~\ref{sec2:fig1} (see \cite{Ramezani.2023,Holmes.2023}) with the use of quantum amplitude amplification (QAA) algorithm \cite{Brassard.2002} to boost the success probability. A sketch is provided in Sect.~\ref{subsec2-1}.  
While the quantum circuit is known, it is not satisfactory for large $N$, especially when the quantum states $\ket{\psi}$ and $\ket{\varphi}$ come from continuous real-valued functions in some applications. In such cases, $\sum_{j=0}^{N-1}|\psi_j\varphi_j|^2=O(1/N)$, and the quantum complexity scales as $O\left(\sqrt{N}\right)$ even if QAA is applied (it scales as $O(N)$ without QAA). 

In this paper, we are concerned with the specific cases that $\ket{\psi}$ and $\ket{\varphi}$ are given by two piecewise continuous real-valued functions $f$ and $g$, respectively. In other words, we assume $f,g: [0, L]\rightarrow \mathbb{R}$ and grid number $N=2^n$, and that the amplitudes of the quantum states satisfy the following relations:
\begin{equation}
\label{sec1:def1}
\psi_j = \frac{f(x_j)}{\sqrt{\sum_{j=0}^{N-1}|f(x_j)|^2}}, \quad \varphi_j = \frac{g(x_j)}{\sqrt{\sum_{j=0}^{N-1}|g(x_j)|^2}}, \quad j=0,\ldots, N-1,
\end{equation}
where $x_j := jL/N$ is the grid point. Under this setting, we propose an $N$-efficient QHP quantum circuit using convolutions in the Fourier domain and show the following theorem.
\begin{theorem}
\label{sec1:thm1}
For two piecewise continuous real-valued functions $f,g$ and arbitrarily given $n\in \mathbb{N}$, assume that the quantum states $\ket{\psi}$, $\ket{\varphi}$ are given by Eq.~\eqref{sec1:def1}. Then, for an error bound $\varepsilon>0$, there exists a quantum circuit to solve the QHP problem with $O\left(1/\varepsilon^s\right)$ queries of the oracles $U_\psi$, $U_\varphi$ and $O\left(n^2/\varepsilon^s\right)$ additional single-qubit or two-qubits gates. Here, $s\ge 0$ is a constant depending on the functions $f$ and $g$, but is independent of $N$.
\end{theorem}
According to the quantum circuit in Sect.~\ref{sec2}, Theorem \ref{sec1:thm1} holds true immediately for multi-variable functions with rectangular domain. Moreover, the above parameter $s$ depends on the smoothness of the functions $f$ and $g$. Smoother functions yield a smaller parameter $s$, and hence, imply a better error bound dependence in the quantum complexity. For piecewise absolutely continuous functions, the parameter $s$ satisfies $s\in [0,1]$. 
In particular, if either $f$ or $g$ has a finite Fourier series expansion, i.e., the number of non-zero Fourier coefficients is an integer that is independent of the above grid number $N$, then we have the following corollary.
\begin{corollary}
\label{sec1:coro1}
Let $f,g$ be two real-valued functions. Assume that either $f$ or $g$ has a finite number of non-zero Fourier coefficients, and the quantum states $\ket{\psi}$, $\ket{\varphi}$ are given by Eq.~\eqref{sec1:def1} for arbitrarily given $n\in \mathbb{N}$. Then, for arbitrary $\varepsilon\ge 0$, there exists a quantum circuit to solve the QHP problem with $O\left(1\right)$ queries of the oracles $U_\psi$, $U_\varphi$ and $O\left(n^2\right)$ additional single-qubit or two-qubits gates.
\end{corollary}

Here, $O(1)$ indicates that the number of queries to the oracles is independent of the grid number $N$ and the error bound error $\varepsilon$. The proofs of Theorem \ref{sec1:thm1} and Corollary \ref{sec1:coro1} immediately follow from the constructed quantum circuit and discussion in Sects.~\ref{subsec2-2}--\ref{subsec2-3}. 
We mention that Proposition \ref{sec1:prop1} and Corollary \ref{sec1:coro1} holds true for arbitrary $\varepsilon \ge 0$. This implies that the QHP state is precisely prepared as we can take $\varepsilon=0$. 

The rest of the paper is organized as follows. In Sect.~\ref{sec2}, we provide the proposed quantum circuit for the QHP problem. The sketch of proofs for the main results are provided by estimating the approximation parameter in the quantum circuit. Sect.~\ref{sec3} is devoted to some applications of the proposed circuit. In particular, an efficient quantum circuit for the partial inner product is further proposed. We end up the paper with a concluding section, and provide further details in the appendix.     

\section{Method}
\label{sec2}

\subsection{Revisit of conventional method}
\label{subsec2-1}

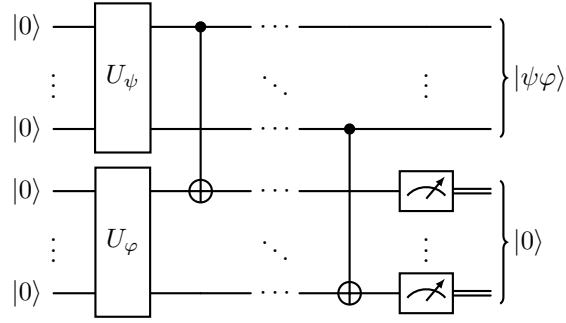
\begin{figure}
\centering
\begin{quantikz}
\lstick{\ket{0}} & \gate[3]{U_\psi} & \ctrl{3} & \ \cdots \ &  &  & \rstick[3]{\ket{\psi\varphi}}\\[-0.4cm]
\setwiretype{n} \vdots &  &  & \ddots &  & \vdots &  & \\[-0.4cm]
\lstick{\ket{0}} &  &  &  \ \cdots \  & \ctrl{3} &  & \\[-0.3cm]
\lstick{\ket{0}} & \gate[3]{U_\varphi} & \targ{} &  \ \cdots \  &  & \meter{} & \setwiretype{c} \rstick[3]{\ket{0}} \\[-0.4cm]
\setwiretype{n} \vdots &  &  & \ddots &  & \vdots &  & \\[-0.4cm]
\lstick{\ket{0}} &  &  &  \ \cdots \  & \targ{} & \meter{} & \setwiretype{c} 
\end{quantikz}
\caption{Conventional quantum circuit for the QHP problem, e.g., \cite{Ramezani.2023,Holmes.2023}. }
\label{sec2:fig1}
\end{figure}

The fundamental method for the QHP problem uses a series of CNOT gates and a post-selection (measurements of partial qubits and selection of desirable results) to link the computational bases of the quantum states $\ket{\psi}$ and $\ket{\varphi}$. The quantum circuit is described in Fig.~\ref{sec2:fig1}. 
It is readily to check that the quantum state before the measurements is as follows:
\begin{align*}
\ket{0}_n \otimes \sum_{j=0}^{N-1} \psi_j\varphi_j \ket{j}_n + \cdots
\end{align*}
Thus, the success probability of post-selecting the significant $n$ qubits as $\ket{0}_n$ can be calculated by
\begin{align*}
\mathbb{P}_{\text{suc}} = \sum_{j=0}^{N-1} |\psi_j\varphi_j|^2.
\end{align*}
Using QAA algorithm, we can define an amplification operator $\mathcal{Q}$, which uses two queries of $U_\psi$ and $U_\varphi$ and at most $O(n^2)$ additional single-qubit or two-qubits gates that come from the reflection operator \cite{Brassard.2002}, and apply it for $O(\sqrt{1/\mathbb{P}_\text{suc}})$ times to boost the success probability to a constant near one. This proves Proposition \ref{sec1:prop1}. 

However, except for the cases that $\ket{\psi}$ or $\ket{\varphi}$ has only several non-zero amplitudes, the above success probability has a scaling of $O(1/N)$ in general. In particular, this occurs when $\ket{\psi}$ and $\ket{\varphi}$ are given by two piecewise continuous real-valued functions as Eq.~\eqref{sec1:def1}. A direct calculation yields
\begin{align*}
\mathbb{P}_{\text{suc}} &= \sum_{j=0}^{N-1} |\psi_j\varphi_j|^2 = \frac{1}{\left(\sum_{j=0}^{N-1}|f(x_j)|^2\right)\left(\sum_{j=0}^{N-1}|g(x_j)|^2\right)} \sum_{j=0}^{N-1} |f(x_j)g(x_j)|^2 \\
&= \frac{L}{N} \frac{\frac{L}{N}\sum_{j=0}^{N-1} |f(x_j)g(x_j)|^2}{\left(\frac{L}{N}\sum_{j=0}^{N-1}|f(x_j)|^2\right)\left(\frac{L}{N}\sum_{j=0}^{N-1}|g(x_j)|^2\right)}
\approx \frac{L}{N} \frac{\|fg\|_{L^2}^2}{\|f\|_{L^2}^2 \|g\|_{L^2}^2} = O(1/N),
\end{align*}
for large grid number $N$. Here, $\|\cdot\|_{L^2}$ denotes the norm of the Lebesgue space $L^2$ \cite{Adams.2003} and is independent of $N$. 
In the next subsection, we propose a new method for the QHP problem with a larger success probability at the cost of possibly introducing some approximation error. 

\subsection{Quantum Hadamard product via Fourier space}
\label{subsec2-2}

The idea is to translate the Hadamard product into the convolution operation in the Fourier space, and to apply approximation for the convolution operation. As we can see in the following context, the convolution operation in the Fourier space itself does not help improve the success probability. 
Whereas, the localized Fourier coefficients enable us to introduce some approximation parameter $M\le N$, and this can enlarge the success probability without additional queries to the oracles $U_\psi$ and $U_\varphi$. 

\begin{figure}
\centering
\begin{quantikz}
\lstick{\ket{0}} & \qwbundle{n} & \gate{U_\psi} & \gate{U_{\text{QFT}}^\dag} & \gate[4]{U_{\text{MADD}}} & \gate{U_{\text{QFT}}} &  &  & \rstick{$\ket{\phi_\text{out}}\approx \ket{\psi\varphi}$}\\
\lstick{\ket{0}} & \qwbundle{\!\!\!m-1} & \gate[3]{U_\varphi} & \gate[3]{U_{\text{QFT}}^\dag} &  &  & \gate[2]{H^{\otimes m}} & \meter{} & \setwiretype{c} \rstick{\ket{0}}\\
\lstick{\ket{0}} & \qw &  &  &  & \gate[2]{U_{+1}} &  & \meter{} & \setwiretype{c} \rstick{\ket{0}}\\
\lstick{\ket{0}} & \qwbundle{\!\!\!n-m} &  &  &  &  &  & \meter{} & \setwiretype{c} \rstick{\ket{0}}
\end{quantikz}
\caption{A novel quantum circuit for the QHP problem. Here, $M=2^m$ is an approximation parameter which is determined in advance. $U_{\text{MADD}}$ is the quantum modular adder \cite[Fig. 9]{Li.2020} and $U_{+1}$ is the quantum incrementer \cite{Gidney2015pre}.}
\label{sec2:fig2}
\end{figure}
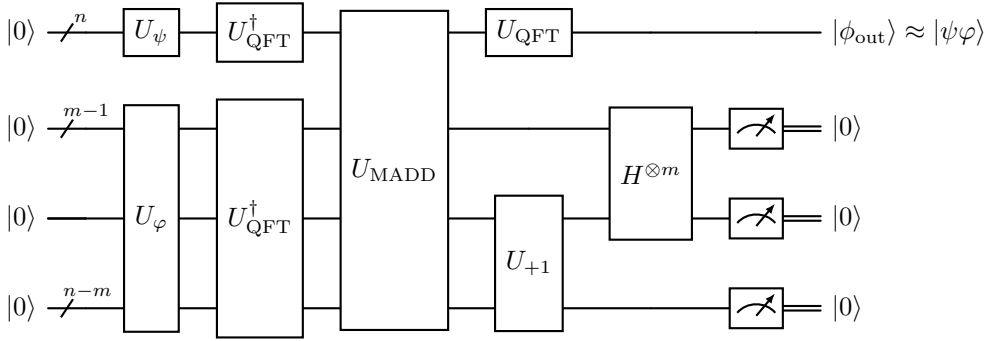

Assume that $M=2^m$ for a positive integer $m\le n$. In this manuscript, we call $M$ the approximation parameter. The proposed quantum circuit for the QHP problem is demonstrated in Fig.~\ref{sec2:fig2}. We describe the operations step by step to show the detailed procedures to prepare the desired quantum state. 

\noindent \underline{Step 1} Apply the oracles $U_\psi$ and $U_\varphi$ to the first and second $n$ qubits, respectively:
$$
(U_{\varphi} \otimes U_{\psi}) (\ket{0}_n \otimes \ket{0}_n) = \sum_{j^\prime=0}^{N-1} \varphi_{j^\prime} \ket{j^\prime}_n \otimes \sum_{j=0}^{N-1} \psi_{j} \ket{j}_n =: \ket{\phi_1}_{2n}.
$$

\noindent \underline{Step 2} Apply the inverse quantum Fourier transforms (QFTs) to the first and second $n$ qubits, respectively to shift to the Fourier bases: 
$$
(U_{\text{QFT}}^\dag \otimes U_{\text{QFT}}^\dag) \ket{\phi_1}_{2n} = \sum_{k^\prime=0}^{N-1} d_{k^\prime} \ket{k^\prime}_n \otimes \sum_{k=0}^{N-1} c_{k} \ket{k}_n =: \ket{\phi_2}_{2n},
$$
where 
$$
c_k = \frac{1}{\sqrt{N}}\sum_{j=0}^{N-1} \psi_j e^{-\mathrm{i}\frac{2\pi}{N} kj}, \quad d_{k^\prime} = \frac{1}{\sqrt{N}}\sum_{j^\prime=0}^{N-1} \varphi_{j^\prime} e^{-\mathrm{i}\frac{2\pi}{N} k^\prime j^\prime}.
$$
As well known, the QFT or its inverse has a total gate count (in basis gates) of $O(n^2)$ and a depth of $O(n)$.  

\noindent \underline{Step 3} Apply the quantum modular adder in the Fourier bases:
$$
U_{\text{MADD}} \ket{\phi_2}_{2n} = \sum_{k,k^\prime=0}^{N-1} c_k d_{k^\prime} \ket{k^\prime}_n \otimes \ket{k+k^\prime}_n =: \ket{\phi_3}_{2n}.
$$
The quantum modular adder proposed in Sect. 4.2 in \cite{Li.2020} has a T-depth of $6n-9$ and a total depth of $16n-19$, which is linear regarding $n$. 

\noindent \underline{Step 4} Apply the QFT to shift the first $n$ qubits back to the original bases:
$$
(I^{\otimes n}\otimes U_{\text{QFT}}) \ket{\phi_3}_{2n} = \frac{1}{\sqrt{N}} \sum_{j=0}^{N-1} \sum_{k,k^\prime=0}^{N-1} c_k d_{k^\prime} e^{\mathrm{i}\frac{2\pi}{N}(k+k^\prime)j} \ket{k^\prime}_n \otimes \ket{j}_n =: \ket{\phi_4}_{2n}.
$$

\noindent \underline{Step 5} Apply the quantum incrementer to the last $n-m+1$ qubits, which is equivalent to a shift of basis by $M/2$ in the last $n$ qubits:
\begin{align*}
(U_{+1} \otimes I^{\otimes n+m-1}) \ket{\phi_4}_{2n} &= \frac{1}{\sqrt{N}} \sum_{j=0}^{N-1} \sum_{k,k^\prime=0}^{N-1} c_k d_{k^\prime} e^{\mathrm{i}\frac{2\pi}{N}(k+k^\prime)j} \ket{k^\prime+M/2}_n \otimes \ket{j}_n \\
&= \frac{1}{\sqrt{N}} \sum_{j=0}^{N-1} \sum_{k,k^{\prime\prime}=0}^{N-1} c_k d_{k^{\prime\prime}-M/2} e^{\mathrm{i}\frac{2\pi}{N}(k+k^{\prime\prime}-M/2)j} \ket{k^{\prime\prime}}_n \otimes \ket{j}_n =: \ket{\phi_5}_{2n}.
\end{align*}
There are several implementations of the quantum incrementer \cite{Gidney2015pre, Yuan.2023}. The well-known one \cite{Gidney2015pre} uses a NOT gate and $n-1$ (multi-)controlled NOT gates. The total gate count is at most $O(n^2)$ without any ancillary qubits, and we obtain a linear order $O(n)$ in both gate count and depth if one ancillary qubit is available. 

\noindent \underline{Step 6} Apply Hadamard gates to the first $m$ qubits in the second part of the quantum circuit, which yields a linear combination of the amplitudes for $\ket{k^{\prime\prime}}_n=\ket{0}_n,\ldots,\ket{M-1}_n$: 
\begin{align*}
&(I^{\otimes n-m} \otimes H^{\otimes m} \otimes I^{\otimes n}) \ket{\phi_5}_{2n} \\
&= \frac{1}{\sqrt{MN}} \sum_{j,k=0}^{N-1} \sum_{k^{\prime\prime}_{1}=0}^{N/M-1}\sum_{k^{\prime\prime}_2, \tilde{k}=0}^{M-1} (-1)^{\sigma(k_2^{\prime\prime}, \tilde{k})} c_k d_{Mk^{\prime\prime}_1+k^{\prime\prime}_2-M/2} e^{\mathrm{i}\frac{2\pi}{N}(k+Mk^{\prime\prime}_1+k^{\prime\prime}_2-M/2)j} \ket{k^{\prime\prime}_1}_{n-m} \otimes \ket{\tilde{k}}_m \otimes \ket{j}_n \\
&= \frac{1}{\sqrt{MN}} \sum_{j,k=0}^{N-1} \sum_{k^{\prime\prime}_2=0}^{M-1} c_k d_{k^{\prime\prime}_2-M/2} e^{\mathrm{i}\frac{2\pi}{N}(k+k^{\prime\prime}_2-M/2)j} \ket{0}_{n} \otimes \ket{j}_n + \cdots \\
&= \sqrt{\frac{N}{M}} \sum_{j=0}^{N-1} \left(\frac{1}{N}\sum_{k=0}^{N-1}\sum_{\bar{k}=-M/2}^{M/2-1} c_k d_{\bar{k}}e^{\mathrm{i}\frac{2\pi}{N}(k+\bar{k})j}\right)\ket{0}_n \otimes \ket{j}_n + \cdots
\end{align*}
Here, we represent $\ket{k^{\prime\prime}}$ into two parts as $\ket{k^{\prime\prime}}_n=\ket{k^{\prime\prime}_1}_{n-m}\otimes \ket{\tilde k}_m$, and $\sigma(k,\ell) := \sum_{j=0}^{m-1} k_j \ell_j$ is the product of the binary codes for integers $k$ and $\ell$. 

\noindent \underline{Step 7} Post-select the last $n$ qubits to be $\ket{0}_n$. This gives the desired quantum state for sufficiently large $M$ by noting
$$
\frac{1}{N}\sum_{k=0}^{N-1}\sum_{\bar{k}=-M/2}^{M/2-1} c_k d_{\bar{k}}e^{\mathrm{i}\frac{2\pi}{N}(k+\bar{k})j} \approx \frac{1}{N} \sum_{k,\bar{k}=0}^{N-1} c_k d_{\bar{k}} e^{\mathrm{i}\frac{2\pi}{N}(k+\bar{k})j} = \psi_j\varphi_j.
$$
The success probability of the post-selection is 
$$
\mathbb{\tilde P}_\text{suc} = \frac{N}{M} \sum_{j=0}^{N-1} \left|\frac{1}{N}\sum_{k=0}^{N-1}\sum_{\bar{k}=-M/2}^{M/2-1} c_k d_{\bar{k}}e^{\mathrm{i}\frac{2\pi}{N}(k+\bar{k})j}\right|^2 \approx \frac{N}{M} \mathbb{P}_{\text{suc}}.
$$

Through the above discussion, we find that the success probability can be improved if the approximation parameter satisfies $M<N$. To ensure that the derived quantum state is close to the desired quantum state $\ket{\psi\varphi}$, we need to choose $M$ sufficiently large. 

\begin{remark}
The quantum circuit in Fig.~\ref{sec2:fig2} can be readily extended to multi-dimensional cases. We apply all the quantum gates except for the oracles $U_f$ and $U_g$ to each dimension. For example, for a two-dimensional case, we need to apply two quantum modular adder $U_{\text{MADD}}$ to the registers regarding $x$ and the registers regarding $y$, respectively.  
\end{remark}
\begin{remark}
To avoid confusion, we emphasize the difference between our proposed method and the quantum circuit in Fig.~10 in \cite{Ramezani.2023}. Although both quantum circuits apply the QFTs, the target in \cite{Ramezani.2023} is the quantum integer multiplication algorithm, and quantum states for two integers are the input states. This is completely different from our quantum circuit for the QHP problem. Moreover, in addition to the QFTs, a series of CNOT gates is adequate for the problem in \cite{Ramezani.2023}, while we need a quantum modular adder to achieve our target of a QHP state.  
\end{remark}

\subsection{Estimation of $M$ for real-valued functions}
\label{subsec2-3}

\begin{figure}[htb]
\centering
\includegraphics[width=13cm, keepaspectratio]{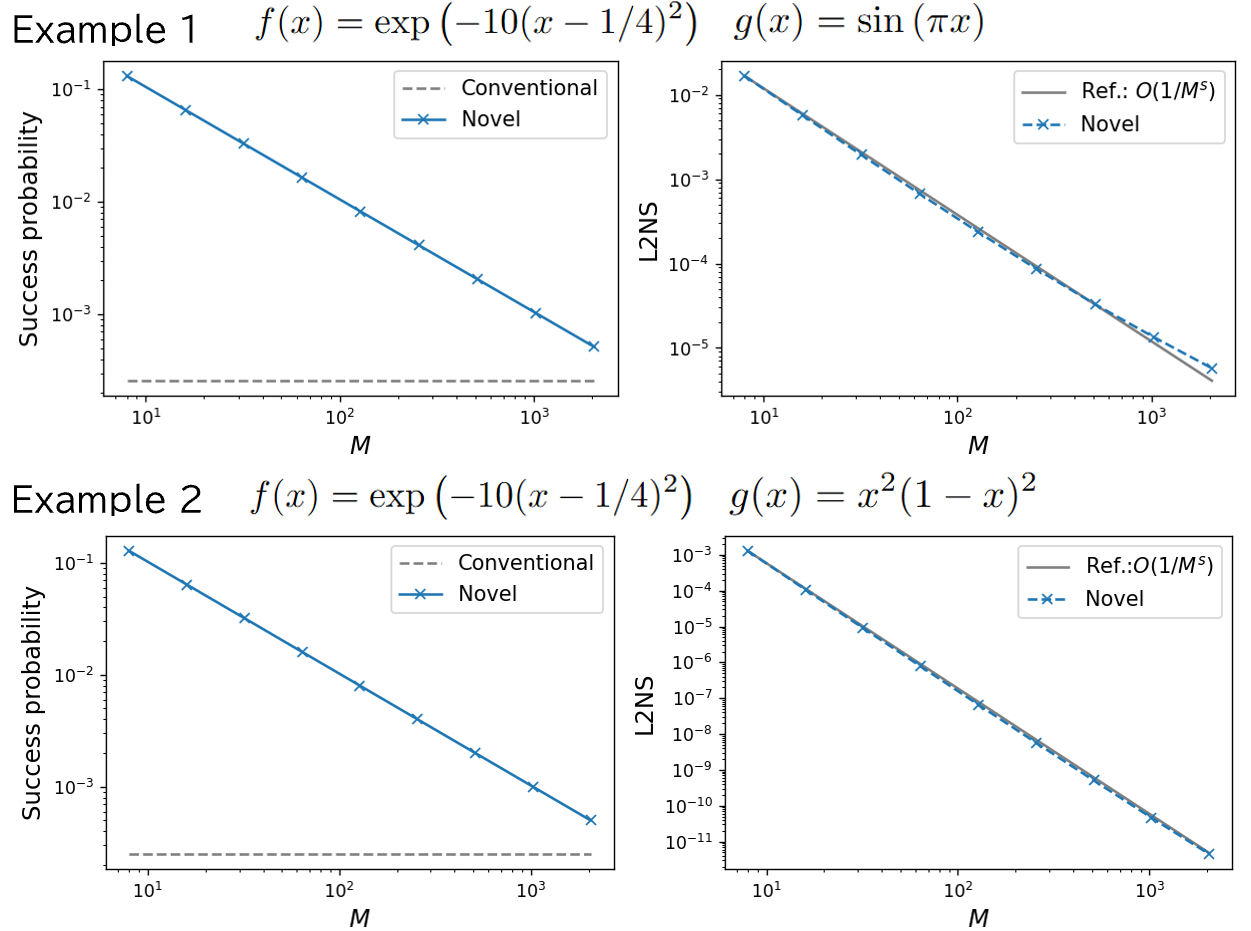} 
\caption{Success probabilities and L2NS errors for Examples 1 and 2 regarding the approximation parameter $M$. For Examples 1 and 2, the gray reference lines denote the orders $O(1/M^{3/2})$ and $O(1/M^{7/2})$, respectively. }
\label{sec2:fig3}
\end{figure}

\begin{figure}[htb]
\centering
\includegraphics[width=13cm, keepaspectratio]{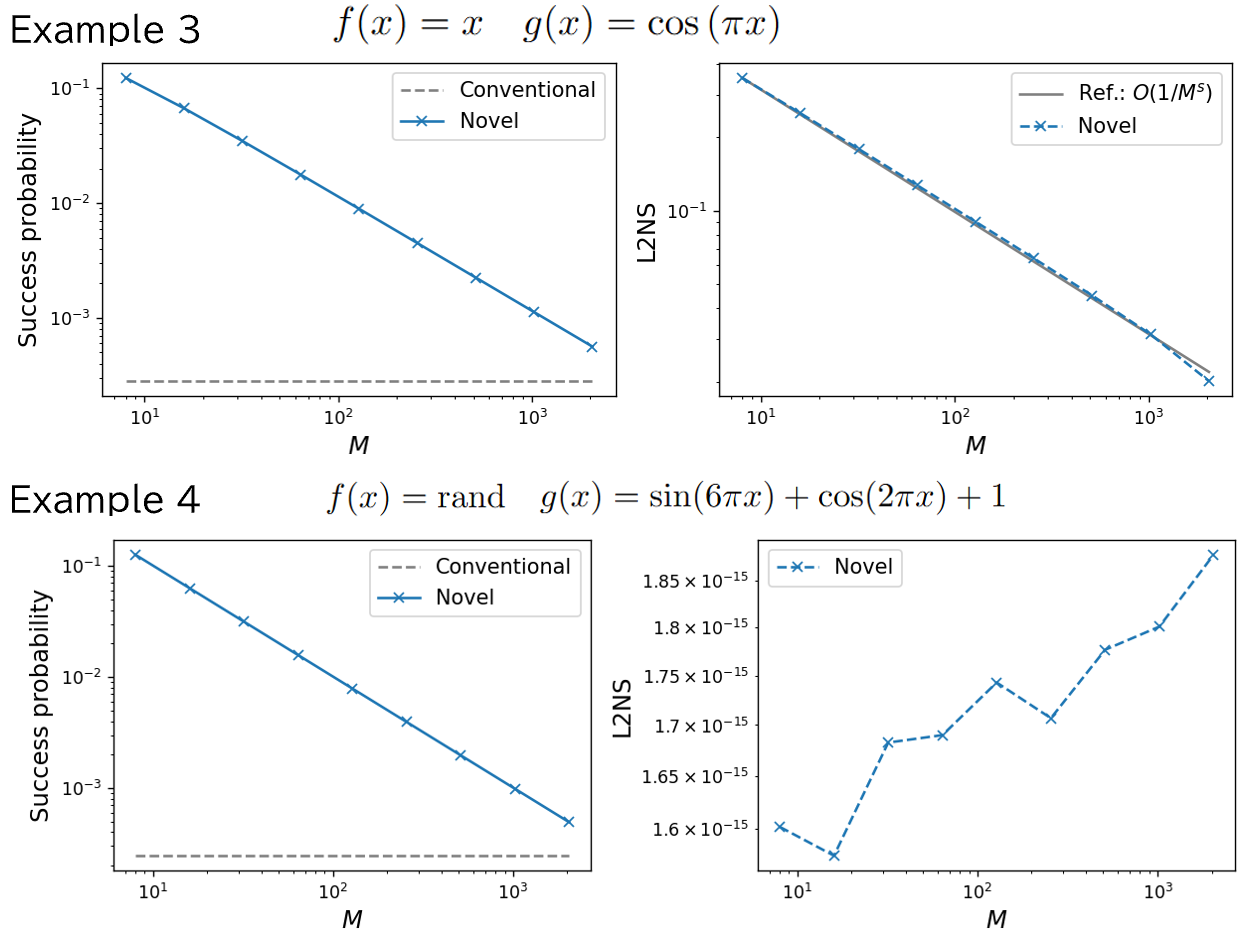} 
\caption{Success probabilities and L2NS errors for Examples 3 and 4 regarding the approximation parameter $M$. For Example 3, the gray reference line denotes the orders $O(1/M^{1/2})$. }
\label{sec2:fig4}
\end{figure}

Although the choice of $M$ is non-trivial in general, we can estimate $M$ in some specific cases, e.g., the quantum states $\ket{\psi}$ and $\ket{\varphi}$ come from some piecewise continuous real-valued functions as in Eq.~\eqref{sec1:def1}. 
Similar to the estimates in \cite{Huang.2025pre}, the approximation parameter $M$ depends on the smoothness of the underlying functions $f,g$ as well as the error bound $\varepsilon$. More precisely, we have following estimate:
\begin{equation}
\label{sec2:eq-M}
M = O((1/\varepsilon)^{s}),
\end{equation}
for some constant $s\ge 0$ which depends on the smoothness of the function $g$. The better localization of the Fourier coefficients gives a smaller $s$, and thus provide a better scaling with respect to the error bound $\varepsilon$. Noting that the quantum circuit in Fig.~\ref{sec2:fig2} is not symmetric regarding $\ket{\psi}$ and $\ket{\varphi}$, it is better to exchange the operations $U_\psi$ and $U_\varphi$ in Fig.~\ref{sec2:fig2} if the smoothness of function $f$ is better than $g$. Therefore, without loss of generality, we can assume that the Fourier coefficients of $g$ decay faster than those of $f$.  
Next, we confirm the parameter $s$ numerically for some elementary functions using the following four examples:
\begin{itemize}
\item \noindent \underline{Example 1} $f(x)=e^{-10(x-1/4)^2}$, $g(x)=\sin(\pi x)$, $x\in [0, 1]$. 

\item \noindent \underline{Example 2} $f(x)=e^{-10(x-1/4)^2}$, $g(x)=x^2(1-x)^2$, $x\in [0, 1]$. 

\item \noindent \underline{Example 3} $f(x)=x$, $g(x)=\cos(\pi x)$, $x\in [0, 1]$. 

\item \noindent \underline{Example 4} $f(x)=\mathrm{arbitrary}$, $g(x)=\sin(6\pi x)+\cos(2\pi x)+1$, $x\in [0, 1]$. 
\end{itemize}
We fix a large grid number $N=2^n=4096$ with $n=12$, and define the $l^2$ error for normalized solutions (L2NS) as follows:
$$
\mathrm{Err}_{\text{L2NS}} := \|\ket{\phi_{\text{out}}}-\ket{\psi\varphi}\|.
$$
To demonstrate the behavior of the proposed quantum circuit by the plots showing both the L2NS errors and the success probabilities regarding several choices of $M=2^3,2^4,\ldots,2^{11}$.  
In Fig.~\ref{sec2:fig3}, Examples 1 and 2 are addressed. According to the error plots, we find $\mathrm{Err}_\text{L2NS} = O(1/M^{\hat s})$ for both examples. For Example 1, we have the decay order $\hat s=3/2$, while we have a larger decay order $\hat s=7/2$ for Example 2. As explained in \cite{Huang.2025pre}, the decay order is $\hat s=(2p-1)/2$ where $p\ge 1$ is the largest integer such that $(\hat p-2)$-th derivative of the function is continuous and $(\hat p-1)$-th derivative is integrable as a periodic function for any $\hat p=1,\ldots,p$. For the above elementary functions, the discontinuities come from the boundary points. For example, we have $g(0) = g(1)$ and $\partial_x g(0) \not= \partial_x g(1)$ for Example 1, and $\partial_x^3 g(0) \not= \partial_x^3 g(1)$ for Example 2. 
On the other hand, Examples 3 and 4 are plotted in Fig.~\ref{sec2:fig4}. Example 3 gives the worst case that the decay order is $s=1/2$ since $g(0) \not= g(1)$, while Example 4 gives the best scenario that $g$ is a smooth periodic function such that $\hat s=\infty$ (only round-off error remains for $M\ge 8$). Moreover, we take a random vector for $f$ in Example 4 to show that the error is mainly affected by the function $g$. 
To sum up, we confirm Eq.~\eqref{sec2:eq-M} with $s=1/\hat s=2/(2p-1)\in [0, 2]$ for $p\in \{1,2,\ldots\}$. In particular, for a smooth periodic function as Example 4, $\hat s=\infty$ implies $s=0$, and this should be explained that the approximation parameter $M$ is independent of the error bound $\varepsilon$. 

Moreover, the plots of success probabilities in Figs.~\ref{sec2:fig3}-\ref{sec2:fig4} implies a clear relation $\mathbb{P}_\text{suc}=O(1/M)$. Together with Eq.~\eqref{sec2:eq-M}, we obtain $\mathbb{P}_\text{suc}=O(\varepsilon^s)$. Therefore, the number of queries of the oracles is $O(1/\mathbb{P}_\text{suc})=O(1/\varepsilon^s)$ for the proposed quantum circuit, and we need at most $O(n^2/\varepsilon^s)$ additional single-qubit or two-qubits gates where $n^2$ comes from the gate counts for the QFTs, quantum modular adders, etc. This proves Theorem \ref{sec1:thm1}. 
Furthermore, similar to Proposition \ref{sec1:prop1}, we can apply the QAA to boost the success probability using a deeper quantum circuit of depth $O((1/\varepsilon)^{s/2})$. Then, the quantum complexity (i.e., the number of queries to the oracles) admits a square root reduction. 
Corollary \ref{sec1:coro1} follows immediately from Theorem \ref{sec1:thm1} since $s=0$ for a function with finite number of Fourier coefficients. 

\begin{figure}[htb]
\centering
\begin{quantikz}
\lstick{\ket{0}} & \qwbundle{n} & \gate{U_\psi} & \gate{U_{\text{QFT}}^\dag} & \gate[2]{U_{\text{MADD}}} & \gate{U_{\text{QFT}}} & \\
\lstick{\ket{0}} & \qwbundle{n} & \gate{U_\varphi} & \gate{U_{\text{QFT}}^\dag} &  & \meter{} & \setwiretype{c} 
\end{quantikz}
\caption{Quantum circuit to determine the approximation parameter $M$.}
\label{sec2:fig5}
\end{figure}
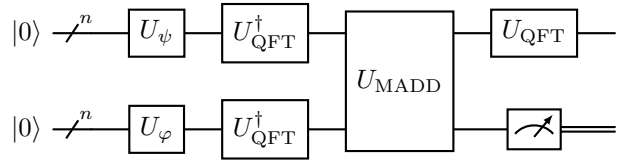

We end up this subsection by a practical choice of the approximation parameter $M$. The above discussion gives only the theoretical order of the approximation parameter regarding the error bound. On the other hand, it is expensive to try several approximation parameters and evaluate the error between the obtained quantum state and the desired quantum state. In practice, we suggest the following way to determine the approximation parameter:
\begin{itemize}
\item Apply the quantum circuit in Fig.~\ref{sec2:fig5} for $N_{\text{shot}}$ times to make a histogram, which indicates the number of counts $N_{\text{count},k}$ for each basis $\ket{k}$, $k=0,1,\ldots,N-1$. Here, $N_\text{shot}$ is a given number in advance, e.g., $N_{\text{shot}}=1000$.

\item Calculate 
$$
C(\ell) := \frac{1}{N_{\text{shot}}}\sum_{k=-2^{\ell-1}}^{2^{\ell-1}-1} N_{\text{count},k},
$$
from $\ell=1$ until the condition $1-C(\ell)\le \epsilon_0$ is satisfied. Here, $\epsilon_0$ is a given error tolerance in advance, e.g., $\epsilon_0=\varepsilon^2$. 

\item Determine the approximation parameter by $M=2^\ell$. 
\end{itemize}
We can choose to ignore the counts $N_{\text{count},k}$ if it is one and its neighbors are zero to avoid the case that a basis with an extremely small amplitude is sampled occasionally. The main additional cost is the $N_{\text{shot}}$ queries of the oracles $U_\psi,U_\varphi$. 

\subsection{Scope of applications of proposed method}

We give several comments on the scope of applications. 

While we mainly consider the applications to real-valued quantum states, the proposed methods can be applied for general quantum states. The key requirement is that the Fourier coefficients are localized, and it is automatically satisfied when we consider real-valued quantum states for suitable real-valued functions.

If the function $g$ (or $f$) satisfies the condition:
$$
g\in W^{2,1}(0,L), \quad g(0)=g(L),
$$
where $[0, L]$ is the domain of the function, and $W^{k,p}$ denotes the Sobolev space \cite{Adams.2003}, then the Fourier coefficients of $g$ have a decay order: $|c_k| = O(k^{-2})$ \cite{Huang.2025pre}. This condition is identical to the fact that the function has the same boundary values, and its first derivative is absolutely continuous. 
In such a case, we find $M=64$ is sufficient to derive a quantum state with an L2NS error smaller than $10^{-3}$, see Examples 1,2,4. 

The proposed method can be readily extended to multi-dimensional cases. The quantum circuit is similar to Fig.~\ref{sec2:fig2}. One difference is that we apply the QFTs, the quantum modular adders, the quantum incrementers and the Hadamard gates to the quantum register for each dimension separately. 

On the other hand, the decay order of Fourier coefficient is not satisfactory when the boundary values does not coincide. One remedy is to use two ancillary qubits to make even extensions of the quantum states. This helps us to retrieve a decay order of $O(k^{-2})$. Whereas, for the discontinuous functions, we arrive at the worst order $|c_k|=O(k^{-1})$, and hence $M=O((1/\varepsilon)^2)$. In addition, the proposed method is not suitable for the quantum state $\ket{\varphi}$ with limited number of non-zero amplitudes. For such a case, the conventional method is better.  

According to the theoretical decay order, the proposed method works well when $g$ (or $f$) is periodic and sufficiently smooth. One natural application is that the quantum state $\ket{\varphi}$ (or $\ket{\psi}$) represents a solution to some partial differential equation (PDE) with periodic boundary condition. 

\section{Applications}
\label{sec3}

We consider the applications of the quantum Hadamard product.

\subsection{Diagonal matrices}

Given a diagonal matrix $D\in \mathbb{R}^{N\times N}$ and a vector $\mathbf{b}\in \mathbb{R}^N$ with the following expressions in quantum fashion:
$$
D = \sum_{j=0}^{N-1} d_j \ket{j}\bra{j}, \quad \mathbf{b} = \sum_{j=0}^{N-1} b_j \ket{j}. 
$$
To compute the matrix-vector multiplication using quantum computing is to prepare a quantum state:
$$
\ket{D\mathbf{b}} := \frac{1}{\sqrt{\sum_{j^\prime=0}^{N-1}|d_{j^\prime}b_{j^\prime}|^2}}\sum_{j=0}^{N-1}d_j b_j \ket{j} .
$$
A conventional way is to prepare a block encoding of the matrix: $U_D$ and apply it to the quantum state corresponding to the vector: $\ket{\mathbf{b}}$:
$$
\ket{D\mathbf{b}} = U_D\ket{\mathbf{b}}.
$$
The existing works assume the following oracle:
$$
\mathcal{O}_D: \ket{i}\ket{0} \to \ket{i}\ket{d_i},
$$
or a direct implementation in \cite{Mottonen.2004} with a gate complexity $O(N)$. 
With the QHP quantum circuit in Sect.~\ref{sec2}, we can realize the block encoding $U_D$ without the above oracle. Instead, we need only quantum state preparation (QSP) circuits of the diagonal components of $D$ as well as the vector $\mathbf{b}$:
$$
U_{\mathbf{d}}: \ket{0} \to \ket{\mathbf{d}}= \frac{1}{\|\mathbf{d}\|}\sum_{j=0}^{N-1} d_j\ket{j},\quad U_{\mathbf{b}}: \ket{0} \to \ket{\mathbf{b}}= \frac{1}{\|\mathbf{b}\|}\sum_{j=0}^{N-1} b_j\ket{j},
$$
by letting $\mathbf{b}$ and $\mathbf{d}$ be the functions $f$ and $g$ (or $g$ and $f$), respectively. 

This implementation of diagonal non-unitary operation is extremely helpful when we do not have access to the classical information of both $\mathbf{b}$ and $D$, but only the quantum states corresponding to the vector and the diagonal of the matrix. 
Another advantage of this implementation is that one can utilize the extensively discussed approaches in the QSP, including the matrix product state (MPS) techniques \cite{Schon.2005,Ran2020,Holmes.2020,Gundlapalli.2022}. 

On the other hand, we mention that the implementation via the QHP is not always efficient since the conventional method leads to a $O(1/\sqrt{N})$ success probability. If we know that the underlying function of $\mathbf{b}$ or $\mathbf{d}$ is sufficiently smooth and nearly periodic in advance, then we can employ the efficient QHP quantum circuit proposed in Sect.~\ref{sec2} to obtain a success probability that is independent of the grid number $N$. 

The diagonal non-unitary operation appears frequently in solving the PDEs for practical applications, for example, the implementation of the reaction term using splitting methods for diffusion-reaction equations. 

\subsection{Partial inner product}

Let $T_1,T_2,L>0$ be given. Assume that $f,g: [T_1,T_2] \times [0,L]\to \mathbb{R}$ are two real-valued functions with multiple variables $x\in [0,L]$ and $t\in [T_1,T_2]$. We consider the calculation of the partial inner product: 
\begin{align*}
\mathcal{I}[f,g](x) := \int_{T_1}^{T_2} f(x,t) g(x,t) \mathrm{d}t, \quad x\in [0,L]. 
\end{align*}
Let $N_t, N_x\in \mathbb{N}$ be two grid numbers, and we consider the uniform divisions: $0=x_0<x_1<\cdots<x_{N_x}=L$, $T_1<t_0<t_1<\cdots<t_{N_t-1}<T_2$, where $x_j = jL/N_x$ and $t_i = T_1 + (T_2-T_1)/(2N_t) + i(T_2-T_1)/N_t$ for $i=0,1,\ldots,N_t-1$, $j=0,1,\ldots,N_x$. Then, the above integral can be approximated by a function $h$ and a corresponding vector $\mathbf{h}\in \mathbb{R}^{N_x}$ whose components satisfy
\begin{align*}
h_j = h(x_j) := \frac{T}{N_t}\sum_{i=0}^{N_t-1} f(x_j, t_i) g(x_j,t_i), \quad j=0,1,\ldots,N_x-1.
\end{align*}
Here and henceforth, $T=T_2-T_1$. Moreover, we assume that $f,g$ are sufficiently smooth and 
$$
g(0,t)=g(L,t)\ \text{ for } t\in [T_1, T_2].
$$ 
Thus, we does not include the right boundary point $x=L$.

To discuss the quantum circuit for this problem, we further assume that $N_x = 2^{n_x}$, $N_t = 2^{n_t}$ and that two oracles for the functions $f,g$ are given:
\begin{align*}
U_f\ket{0}_{n_t}\ket{0}_{n_x} = \sum_{i=0}^{N_t-1}\sum_{j=0}^{N_x-1} \tilde{f}_{i,j} \ket{i}_{n_t}\ket{j}_{n_x}, \quad 
U_g\ket{0}_{n_t}\ket{0}_{n_x} = \sum_{i=0}^{N_t-1}\sum_{j=0}^{N_x-1} \tilde{g}_{i,j} \ket{i}_{n_t}\ket{j}_{n_x},
\end{align*}
where 
$$
\tilde{f}_{i,j} = \frac{f(x_j,t_i)}{\sqrt{\sum_{\bar i=0}^{N_t-1}\sum_{\bar j=0}^{N_x-1}|f(x_{\bar j},t_{\bar i})|^2}}=: \frac{f(x_j,t_i)}{A_f},  \quad 
\tilde{g}_{i,j} = \frac{g(x_j,t_i)}{\sqrt{\sum_{\bar i=0}^{N_t-1}\sum_{\bar j=0}^{N_x-1}|g(x_{\bar j},t_{\bar i})|^2}}=:\frac{g(x_j,t_i)}{A_g},
$$ 
for $i=0,1,\ldots,N_t-1$, $j=0,1,\ldots,N_x-1$. Combining the quantum strategy for summation and the QHP quantum circuit in Sect.~\ref{sec2}, we propose the quantum circuit to prepare a quantum state:
$$
\ket{\phi}_{\text{out}} \approx \ket{\mathbf{h}}_{n_x} = \sum_{j=0}^{N_x-1} \tilde{h}_j \ket{j}_{n_x}, 
$$
which is shown in Fig.~\ref{sec3:fig1}. Here, $\tilde{h}_j$ is proportional to $h_j$ by a constant $\sqrt{\sum_{j=0}^{N_x-1}|h(x_j)|^2}$. This gives the approximate quantum state for the partial inner product for two multi-variable functions. 
\begin{figure}[htb]
\centering
\begin{quantikz}[transparent]
\lstick{\ket{0}} & \qwbundle{n_x} & \gate[2]{U_f} & \gate{U_{\text{QFT}}^\dag} & \gate[4]{U_{\text{MADD}}^\dag} & \gate{U_{\text{QFT}}} &  &  & \rstick{\ket{\phi_{\text{out}}}}\\
\lstick{\ket{0}} & \qwbundle{n_t} &  &  & \linethrough &  & \gate[3]{U_g^\dag} & \meter{} & \setwiretype{c} \rstick{\ket{0}}\\
\lstick{\ket{0}} & \qwbundle{m} & \gate{H^{\otimes m}} & \gate[2]{U_{+M/2}^\dag} &  & \gate[2]{U_{\text{QFT}}} &  & \meter{} & \setwiretype{c} \rstick{\ket{0}}\\
\lstick{\ket{0}} & \qwbundle{n_x-m} &  &  &  &  &  & \meter{} & \setwiretype{c} \rstick{\ket{0}}
\end{quantikz}
\caption{A novel quantum circuit for calculating the partial integral for two functions. Here, $M\le N_x$ is a given approximation parameter and $m=\log_2 M$. The last $n_t+n_x$ qubits are post-selected to be $\ket{0}$.
}
\label{sec3:fig1}
\end{figure}
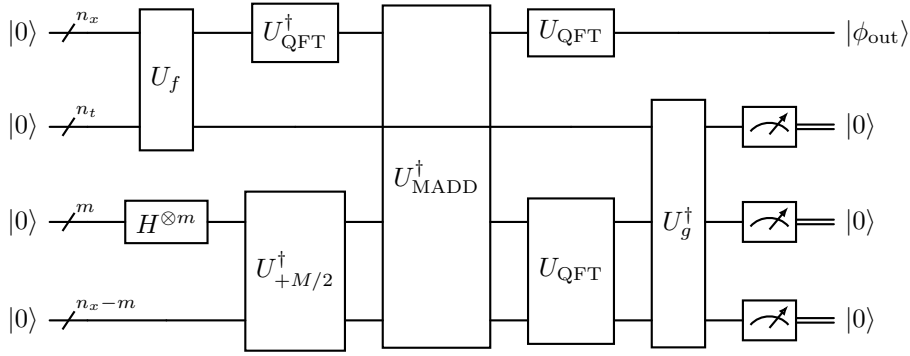
\begin{figure}[htb]
\centering
\begin{quantikz}
\lstick{\ket{0}} & \qwbundle{n_x} & \gate[2]{U_f} &  &  & \ctrl{2} &  &  & \rstick{\ket{\phi_{\text{out}}}}\\
\lstick{\ket{0}} & \qwbundle{n_t} &  &  & \ctrl{2} &  & \gate{H^{\otimes n_t}} & \meter{} & \setwiretype{c} \rstick{\ket{0}} \\
\lstick{\ket{0}} & \qwbundle{n_x} & \gate[2]{U_g} &  &  & \targ{} &  & \meter{} & \setwiretype{c} \rstick{\ket{0}}\\
\lstick{\ket{0}} & \qwbundle{n_t} &  &  & \targ{} &  &  & \meter{} & \setwiretype{c} \rstick{\ket{0}}
\end{quantikz}
\caption{A naive implementation of the partial inner product for two functions. The last $2n_t+n_x$ qubits are post-selected to be $\ket{0}$. The notation of the CNOT gates indicates a series of CNOT gates applied on the corresponding $x$ registers and $t$ registers, respectively.}
\label{sec3:fig2}
\end{figure}
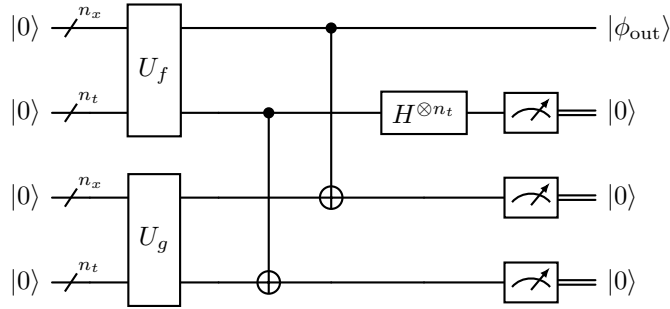
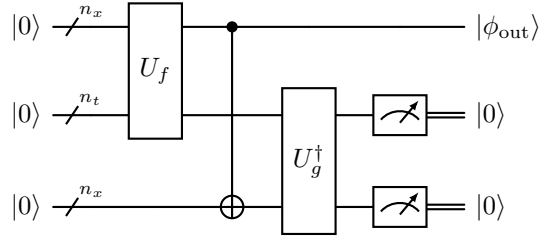
\begin{figure}[htb]
\centering
\begin{quantikz}
\lstick{\ket{0}} & \qwbundle{n_x} & \gate[2]{U_f} & \ctrl{2} &  &  & \rstick{\ket{\phi_{\text{out}}}}\\
\lstick{\ket{0}} & \qwbundle{n_t} &  &  & \gate[2]{U_g^\dag} & \meter{} & \setwiretype{c} \rstick{\ket{0}} \\
\lstick{\ket{0}} & \qwbundle{n_x} &  & \targ{} &  & \meter{} & \setwiretype{c} \rstick{\ket{0}}
\end{quantikz}
\caption{An improved implementation of the partial inner product for two functions. The last $n_t+n_x$ qubits are post-selected to be $\ket{0}$. The notation of the CNOT gate indicates a series of CNOT gates applied on the corresponding $x$ registers.}
\label{sec3:fig3}
\end{figure}
Compared to the naive quantum circuit for the partial inner product in Fig.~\ref{sec3:fig2}, the success probability is improved from $O(1/(N_tN_x))$ to $O(1/M)$, see Table \ref{sec3:tab1}.
\begin{table}[tb]
\centering
\caption{Comparison of different quantum circuit for preparing the partial inner product state. }
\label{sec3:tab1}
\scalebox{1.05}[1.05]{
\begin{tabular}{l|cccc}
\hline
Method & Exact/Approximate & Qubit count & Success probability \\ 
\hline
Conventional (Fig.~\ref{sec3:fig2}) & Exact & $2n_t+2n_x$ & $O\left(1/(N_tN_x)\right)$ \\
Improved (Fig.~\ref{sec3:fig3}) & Exact & $n_t+2n_x$ & $O\left(1/N_x\right)$ \\
Proposed (Fig.~\ref{sec3:fig1}) & Approximate & $n_t+2n_x$ & $O\left(1/M\right)^{\ast}$ \\
\hline
\end{tabular}
}

\vspace{0.1cm}
{\footnotesize\raggedright
$^{\ast}$ According to the discussions in Sect.~\ref{sec2}, $M=O(1/\varepsilon^{s})$ for some $s\ge 0$ where $\varepsilon$ is the error bound. In particular, the proposed method becomes an exact one if $M=N_x$.\\ 
}
\end{table}
One of the contributions is that we extend the inner product technique for calculating a scalar to the partial inner product for calculating a function using only one quantum register for the integration variable, which removes the dependence on the grid number $N_t$, see Fig.~\ref{sec3:fig3}. To the authors' best knowledge, this improved quantum circuit is not known in spite of its simplicity. 
The other contribution is the utilization of the QHP quantum circuit proposed in Sect.~\ref{sec2}, so that we remove the dependence on the grid number $N_x$ provided that the function $g$ is sufficiently smooth and satisfies the periodic boundary condition. 

The partial inner product of two functions has wide applications, including the calculations of some average quantities in certain direction, the integration of the propagators in self-consistent field theory (SCFT) for polymers \cite{Morita.2001, Arora.2016, Ackerman.2017}, etc. 
As for the SCFT calculation, $T_1=0$, and $T_2=T$ denotes the length of a chain. In fact, calculation of a kind of convolution is needed, that is, $f(x, t)$ should be replaced by $f(x, T-t)$. In such a case, we only need to add a series of $X$ gates in the ``time" register after the operation of $U_f$ in the quantum circuit in Fig.~\ref{sec3:fig1}. In other word, the calculation of the following convolution:
\begin{align*}
\mathcal{\tilde I}[f,g](x) := \int_{0}^{T} f(x,T-t) g(x,t) \mathrm{d}t = \int_{0}^{T} \tilde f(x,t) g(x,t) \mathrm{d}t = \mathcal{I}[\tilde f,g](x),  
\end{align*}
can be easily implemented with our proposed methods by denoting $\tilde f(x,t) = f(x,T-t)$. The efficient quantum circuit for the partial inner product enables us to avoid the quantum state tomography of the propagator since only the volume fractions need to be known to update the potential functions for the next SCFT loop. The main quantum circuits for the SCFT calculation for polymers are provided in Appendix A. 

\begin{remark}
To avoid misunderstanding, we emphasize that the proposed quantum circuits in Figs.~\ref{sec3:fig1},\ref{sec3:fig3} actually prepare the quantum state corresponding to 
$$
\int_{T_1}^{T_2} f(x,t)\bar g(x,t) \mathrm{d} t,
$$
where $\bar{\cdot}$ denotes the complex conjugate. This is identical to the desired quantum state assuming that the function $g$ is real-valued. 
\end{remark}

\section{Conclusions}

We propose a new quantum circuit via Fourier space for the quantum Hadamard product (QHP) problem. For two quantum states corresponding to real-valued functions, the proposed method yields a success probability that is independent of the grid number at the expense of approximation error, which is an improvement compared to the conventional QHP circuit. 
In particular, if at least one of the functions has finite number of non-zero Fourier coefficients, then the proposed method improves the success probability without any loss of precision. The proposed quantum circuit is stated for single variable functions, while 

Moreover, the proposed method has wide applications. One typical application is the block encoding of a diagonal matrix. Utilizing the proposed method, we can construct the block encoding of the diagonal matrix by the quantum state preparation of its diagonal components. This is extremely useful when the quantum state of the diagonal instead of its classical information is given, for example, in the quantum algorithm for solving coupling partial difference equations where the interaction appears in the potential term (i.e., zeroth order term). 

Another application is the calculation of the partial inner product of two functions. A discrete use case is the integration of the propagators in self-consistent field theory (SCFT) for polymers. Combining the efficient inner product quantum circuit with the proposed QHP quantum circuit via the Fourier space, we further propose a novel method for such partial inner product calculations. More precisely, we greatly improve the $O(1/N_t N_x)$ success probability of the naive method to a success probability that is independent of both $N_t$ and $N_x$. Here, $N_t$ is the grid number of the integration variable, and $N_x$ is the grid number of rest variables for point-wise calculations. 

\section*{Acknowledgments}
This work was partially supported by the Center of Innovations for Sustainable Quantum AI (JST
Grant number JPMJPF2221).


\section*{Appendix A: Detailed quantum circuit for the convolution operation appeared in the SCFT calculation}

\begin{figure}[htb]
\centering
\resizebox{15cm}{!}{
\begin{quantikz}
\lstick{\ket{q_0}} & \qwbundle{n} & \gate[2]{U_{\text{PITE}}(\Delta t/2)} &  &  & \gate[2]{U_{\text{PITE}}(\Delta t)} &  &  & \gate[2]{U_{\text{PITE}}(2\Delta t)} &  &  & \ \cdots \ & \gate[2]{U_{\text{PITE}}(2^{n_t-1}\Delta t)} &  &  &\rstick[6]{\ket{q}} \\
\lstick{\ket{0}} &  &  & \meter{} & {\setwiretype{c}\ket{0}} & \setwiretype{q} & \meter{} & {\setwiretype{c}\ket{0}} & \setwiretype{q} & \meter{} & {\setwiretype{c}\ket{0}} &  \ \cdots \ \setwiretype{q} &  & \meter{} & {\setwiretype{c}\ket{0}} \\
\lstick{\ket{0}} &  & \gate{H} &  &  & \ctrl{-1} &  &  &  &  &  & \ \cdots \ &  &  &  &\\
\lstick{\ket{0}} &  & \gate{H} &  &  &  &  &  & \ctrl{-2} &  &  & \ \cdots \ &  &  &  &\\
\setwiretype{n} &  & \vdots &  &  &  &  &  &  &  &  & \ \ddots \ &  &  &\\
\lstick{\ket{0}} &  & \gate{H} &  &  &  &  &  &  &  &  & \ \cdots \ & \ctrl{-4} &  &  &
\end{quantikz}
}
\caption{Quantum circuit for preparing the propagator by solving the MDE. }
\label{appA:fig1}
\end{figure}
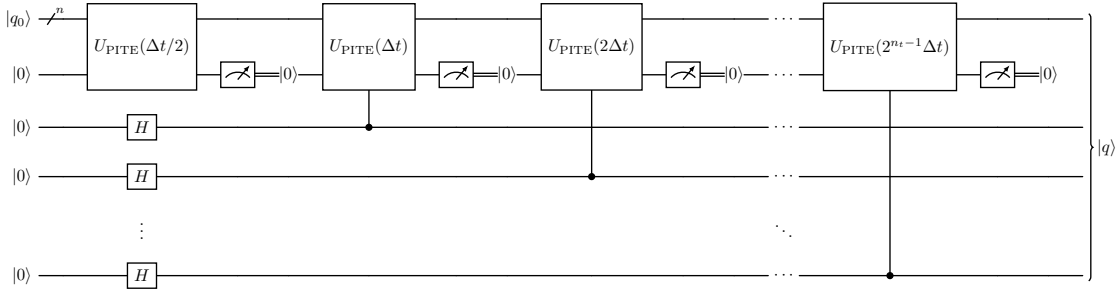

\begin{figure}[htb]
\centering
\begin{quantikz}[transparent]
\lstick{\ket{0}} & \qwbundle{n_x} & \gate[3]{U_q} & \gate{U_{\text{QFT}}^\dag} & \gate[6]{U_{\text{MADD}}^\dag} &  & \gate{U_{\text{QFT}}} &  &  & \rstick[2]{\ket{\phi_\text{out}}}\\[-0.4cm]
\lstick{\ket{0}} & \qwbundle{n_y} &  & \gate{U_{\text{QFT}}^\dag} & \linethrough & \gate[8,label style={yshift=0.4cm}]{U_{\text{MADD}}^\dag} & \gate{U_{\text{QFT}}} &  &  & \\
\lstick{\ket{0}} & \qwbundle{n_t} &  & \gate{X^{\otimes n_t}} & \linethrough & \linethrough &  & \gate[7]{U_q^\dag} & \meter{} & \setwiretype{c} \rstick{\ket{0}}\\
\lstick{\ket{0}} & \qwbundle{\!\!\!\!\!m_x-1} & \gate[2]{H^{\otimes m_x}} &  &  & \linethrough & \gate[3]{U_{\text{QFT}}} &  & \meter{} & \setwiretype{c} \rstick{\ket{0}}\\[-0.4cm]
\lstick{\ket{0}} & \qw{} &  & \gate[2]{U_{+1}^\dag} &  & \linethrough &  &  & \meter{} & \setwiretype{c} \rstick{\ket{0}}\\[-0.4cm]
\lstick{\ket{0}} & \qwbundle{\!\!\!\!\!\!\!n_x-m_x} &  &  &  & \linethrough &  &  & \meter{} & \setwiretype{c} \rstick{\ket{0}}\\[-0.4cm]
\lstick{\ket{0}} & \qwbundle{\!\!\!\!\!m_y-1} & \gate[2]{H^{\otimes m_y}} &  &  &  & \gate[3]{U_{\text{QFT}}} &  & \meter{} & \setwiretype{c} \rstick{\ket{0}}\\[-0.4cm]
\lstick{\ket{0}} & \qw{} &  & \gate[2]{U_{+1}^\dag} &  &  &  &  & \meter{} & \setwiretype{c} \rstick{\ket{0}}\\[-0.4cm]
\lstick{\ket{0}} & \qwbundle{\!\!\!\!\!\!\!n_y-m_y} &  &  &  &  &  &  & \meter{} & \setwiretype{c} \rstick{\ket{0}}
\end{quantikz}
\caption{Quantum circuit for calculating the convolution operation in the SCFT loop for polymers. Here, the ancillary qubits of the oracles $U_q,U_q^\dag$ are omitted. }
\label{appA:fig2}
\end{figure}
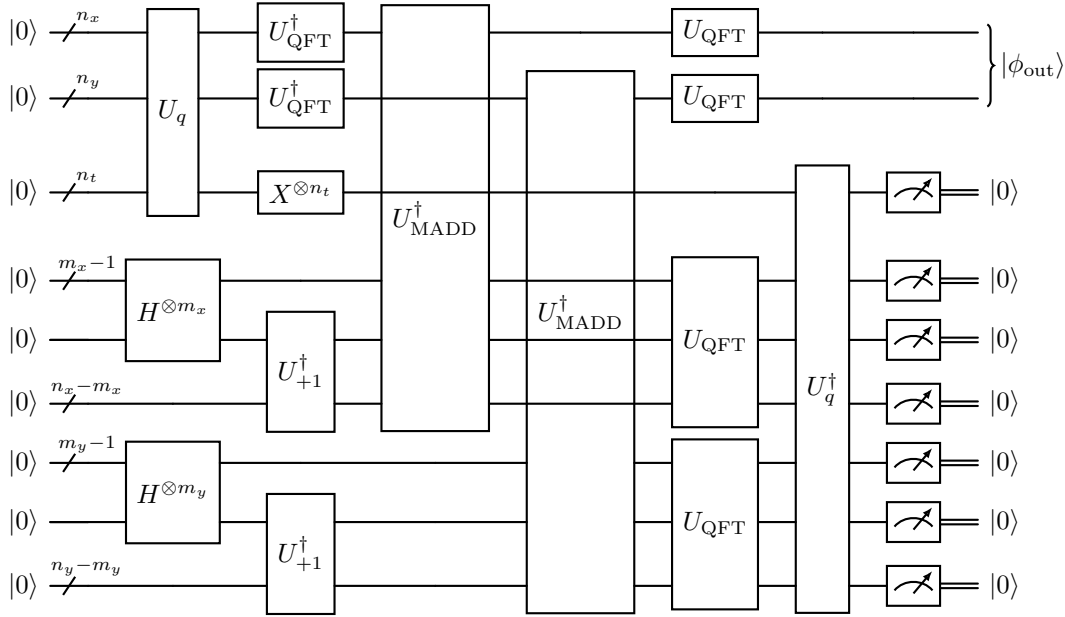

The quantum state of the propagator $q$, the solution of the modified diffusion equation (MDE), can be prepared using the probabilistic imaginary-time evolution (PITE) algorithm \cite{Huang.2024pre} as we show in Fig.~{\ref{appA:fig1}}.
In other words, we have constructed the oracle:
$$
U_q (\ket{0}_{n_t} \otimes \ket{0} \otimes \ket{q_0}_n) = \sum_{j=0}^{K-1} \ket{j}_{n_t} \otimes \ket{0} \otimes \ket{q(\Delta t/2 + j\Delta t)}_n.
$$
Applying the oracle and its inverse, we demonstrate the quantum circuit in Fig.~\ref{appA:fig2} for preparing the quantum state after the convolution operation:
$$
\int_{0}^1 q(s,\mathbf{x}) q(1-s, \mathbf{x}) \mathrm{d} s. 
$$

\end{document}